\documentclass[sigconf]{acmart}

\usepackage{subfig}
\usepackage{booktabs}
\usepackage{multirow}
\usepackage{amsmath}
\usepackage{amsfonts}
\usepackage{algorithm}
\usepackage{algpseudocode}
\usepackage{lipsum}  
\usepackage{bm}
\usepackage{soul}
\usepackage{xcolor}
\usepackage[normalem]{ulem}
\usepackage{xifthen}
\usepackage[switch]{lineno}
\usepackage{pifont}
\usepackage{makecell}
\usepackage{tabularx}

\usepackage{footnote}
\makesavenoteenv{tabular}
\makesavenoteenv{table}
\useunder{\uline}{\ul}{}
\usepackage{makecell}

\newcommand{\longname}{\textit{\uline{P}opularity-\uline{A}ware \uline{Re}commender}}
\newcommand{\shortname}{\textit{PARE}}

\AtBeginDocument{%
  \providecommand\BibTeX{{%
    \normalfont B\kern-0.5em{\scshape i\kern-0.25em b}\kern-0.8em\TeX}}}

\begin{document}

\title{Capturing Popularity Trends: A Simplistic Non-Personalized Approach for Enhanced Item Recommendation}

% \author{Jiazheng Jing$^\dag$, Yinan Zhang$^\dag$, Xin Zhou, Zhiqi Shen$^{*}$}\thanks{$^\dag$ These two authors contributed equally to this work.} \thanks{$^*$ Corresponding author.}
% \affiliation{
% \institution{Nanyang Technological University}
% \country{Singapore}
% }
% \email{
% jiazheng001@e.ntu.edu.sg, {yinan.zhang, xin.zhou, zqshen}@ntu.edu.sg
% }
% \iffalse
\author{Jiazheng Jing}
\authornote{Both authors contributed equally to this research.}
% \author{Yinan Zhang}
% \email{yinan.zhang@ntu.edu.sg}
\affiliation{%
  \institution{Nanyang Technological University}
  % \streetaddress{P.O. Box 1212}
  \country{Singapore}
}
\email{jiazheng001@e.ntu.edu.sg}

\author{Yinan Zhang}
\authornotemark[1]
\affiliation{%
  \institution{Nanyang Technological University}
  % \streetaddress{1 Th{\o}rv{\"a}ld Circle}
  % \city{Hekla}
  \country{Singapore}}
\email{yinan.zhang@ntu.edu.sg}

\author{Xin Zhou}
\affiliation{%
  \institution{Nanyang Technological University}
  \country{Singapore}
}
\email{xin.zhou@ntu.edu.sg}

\author{Zhiqi Shen}
\authornote{Corresponding author.}
\affiliation{%
 \institution{Nanyang Technological University}
 \country{Singapore}}
 \email{zqshen@ntu.edu.sg}

% \affiliation{%

% \affiliation{%

% \affiliation{%

% \affiliation{%
% \fi

\begin{abstract}
  Recommender systems have been gaining increasing research attention over the years. Most existing recommendation methods focus on capturing users' personalized preferences through historical user-item interactions, which may potentially violate user privacy. Additionally, these approaches often overlook the significance of the temporal fluctuation in item popularity that can sway users' decision-making.
  To bridge this gap, we propose \longname{} (\shortname{}), which makes non-personalized recommendations by predicting the items that will attain the highest popularity. 
  \shortname{} consists of four modules, each focusing on a different aspect: popularity history, temporal impact, periodic impact, and side information. Finally, an attention layer is leveraged to fuse the outputs of four modules.
  To our knowledge, this is the first work to explicitly model item popularity in recommendation systems. Extensive experiments show that \shortname{} performs on par or even better than sophisticated state-of-the-art recommendation methods. Since \shortname{} prioritizes item popularity over personalized user preferences, it can enhance existing recommendation methods as a complementary component. Our experiments demonstrate that integrating \shortname{} with existing recommendation methods significantly surpasses the performance of standalone models, highlighting \shortname{}'s potential as a complement to existing recommendation methods. Furthermore, the simplicity of \shortname{} makes it immensely practical for industrial applications and a valuable baseline for future research.
  
\end{abstract}




\maketitle

\section{Introduction}

In recent years, recommendation systems have experienced substantial growth, with applications spanning diverse scenarios such as e-commerce~\cite{Alamdari2020,Yulong2020_eco}, education~\cite{MoodleREC2020,Salazar2021_edu}, and social media~\cite{Aminu2020_social,Hanjraw2019_social}. 
Most existing research works are based on collaborative filtering with the assumption that similar users may interact with similar items~\cite{ncf2017,Wu2023_cf,Xiang2019_ngcf, zhou2023selfcf}.
More specifically, these works make personalized recommendations by leveraging users' historical interaction data to discern individual preferences.
Later, sequential recommendation systems are proposed due to the inherent variations in user preferences, together with the sequential dependencies between their interactions~\cite{Xu2018_seq,tang2018personalized,sun2019bert4rec,Qiaoyu2021_seq,Huang2018_seq}.
% The inherent variations in user preferences, together with the sequential dependencies between their interactions, have led to the proposition of sequential recommendation systems~\cite{Xu2018_seq,tang2018personalized,sun2019bert4rec,Qiaoyu2021_seq,Huang2018_seq}. 
These methods take into account the chronological dynamics of user activities by applying tailored significance factors according to the corresponding interaction timestamps.
However, sequential recommendation methods predominantly target the dynamic nature of user preferences while ignoring the temporal fluctuations in item popularity. 

Predicting item popularity is crucial in enhancing recommendation accuracy and enriching user interaction experiences for several reasons. First, due to a pervasive herd mentality among users ~\cite{Loxton2020_herd,Kurdi2021_herd,Kameda2015_herd}, their decisions are strongly swayed by items' popularity at any given moment. For example, \citet{Robert2023_herd} emphasized that an individual's proclivity towards smoking is strongly influenced by the prevailing smoking rates among his peers.
Besides, findings from \citet{Yitong2020_pop} demonstrated that recommending the most popular movies from the past month significantly outperformed recommending items with the highest global popularity, thereby underscoring the importance of recent item popularity in enhancing recommendation accuracy.
Such effect is particularly apparent for time-sensitive or frequently updated items such as fashionable clothing, movies, and news~\cite{zhang2013_fashion,Miaomiao2019_herd,Li2022_herd}. These domains are particularly susceptible to fluctuations in popularity, necessitating effective prediction strategies.

Second, making recommendations by leveraging item popularity predictions can protect user privacy. 
Concerns have been raised regarding platforms exploiting user interaction history for personalized recommendations, which may compromise user privacy~\cite{Lee2011_privacy,Catherine2014_privacy, al2019privacy}. However, predicting item popularity does not necessarily knowing the precise items that users have interacted with, thus offering a degree of privacy protection.

Third, making recommendations considering the forthcoming item popularity can help mitigate the popularity bias~\cite{Jiawei2023_bias,Himan2020_bias} in recommendation systems to a significant extent.
On the one hand, most existing recommendation methods, which are based on historical user-item interactions, often overlook recommending long-tail or newly released items, given their sparse interaction history. On the other hand, ``classic'' items with a history of high popularity are frequently over-recommended.
% , despite potential interactions on other platforms.
Both cold-start and debiased recommendation methods are proposed to address these challenges~\cite{Yujia2021_cold,Barkan2019_cold,Xinyang2019_bias,Krishnan2018_bias,Ziwei2021_bias}. 
Nevertheless, many cold-start methods capitalize on item properties, leveraging similarity between new releases and previously seen items~\cite{Saveski2014_cold,Xiaoyu2020_cold}, which may lead to unfair treatment for truly novel items.
As for debiased recommendation approaches, some make attempts to uniformly boost the visibility of less-popular items ~\cite{Xinyang2019_bias,Krishnan2018_bias,Abdollahpouri2019_bias}, or employ regularizers to rectify popularity bias~\cite{Ziwei2021_bias,kamishima2014correcting,Zhihong2020_bias}.
However, we believe that a more effective solution lies in accurately predicting future popularity trends.
Such prediction can help surface long-tail items or newly-released items that might be on the cusp of becoming popular, improving the visibility of these less-known items. 
Similarly, items that have already gained popularity can receive the attention they deserve, thus contributing to a fairer and more diverse item recommendation.
% , which allows the promotion of newly released products, as they often attract the most attention upon their initial release.
% popular --> less popular?

\begin{figure}
	\centering
    
	\subfloat[]{\includegraphics[width=.5\columnwidth]{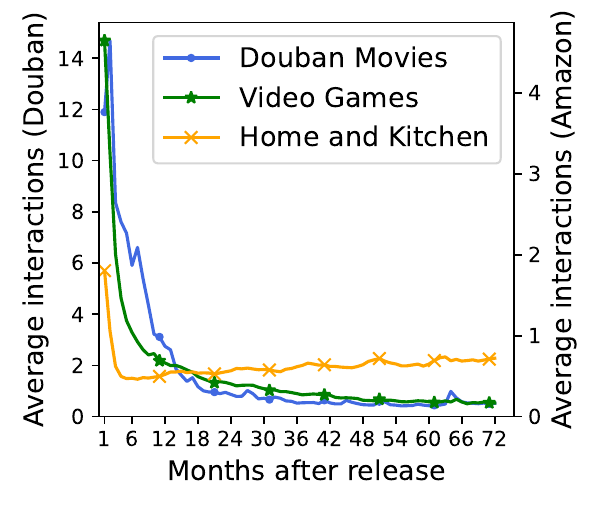}\label{fig:data_all}}
	\subfloat[]{\includegraphics[width=.5\columnwidth]{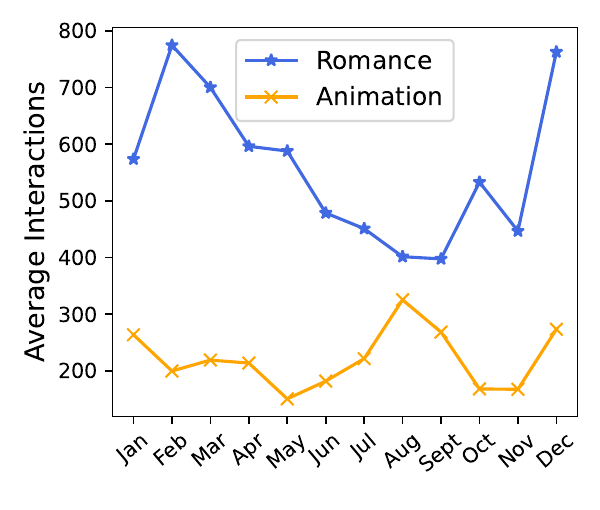}\label{fig:data_genre}}
	\caption{(a) Average monthly interactions of items after being released on three datasets. \textit{Douban Movies} is from Douban. Both \textit{Home and Kitchen} and \textit{Video Games} are from Amazon. (b) Average monthly interactions for movies within the \textit{Romance} and \textit{Animation} genres on \textit{Douban Movies}.}
	\label{fig:dataset_statistic}
\end{figure}

There are many factors that we can take into account when predicting item popularity.
First, the lifecycle of most items typically features periods of prosperity followed by decline. This phenomenon has been illustrated through our empirical analysis of three real-world datasets. Figure \ref{fig:data_all} shows how the average number of interactions on items changes with the time after being released. \textit{Douban Movies} is from Douban\footnote{https://movie.douban.com\label{douban}}, and both \textit{Home and Kitchen} and \textit{Video Games} are from Amazon\footnote{\label{amazon}https://www.amazon.com}. We observe all items tend to attract peak attention within the first two months post-release before rapidly diminishing in popularity, which is even more evident on \textit{Douban Movies} and \textit{Video Games}. We also notice the slowly rising popularity trend for items on \textit{Home and Kitchen} after 1 year of being released. This may be due to the limited product lifespan and the consumers' repurchase.

Moreover, different categories of items undergo various periodic shifts in popularity. As shown in Figure \ref{fig:data_genre}, we analyze the average monthly interactions for movies within the \textit{Romance} and \textit{Animation} genres on \textit{Douban Movies}. Peak attention for \textit{Romance} is observed in February and December, which could be attributed to Valentine's Day and Christmas respectively, both occasions when romantic films are traditionally favored. On the other hand, \textit{Animation} is most popular in August and January, likely coinciding with summer and winter school holidays, during which teenagers tend to have more leisure time.

% \hl{ ADD one more example here, periodic fluctuations in popularity experienced by different item categories, example}

% The analysis underscores the inherent volatility in item popularity, with notable implications for items like movies and video games where popularity typically peaks within the first two months post-release. \hl{under such a phenomenon, cutoff toppop}

Besides categories, other side information may also contribute to the items' popularity. Take movie recommendation as an example, the reputation of the director or the presence of high-profile actors could enhance a movie's attractiveness. 
And user reviews offer crucial insights into the public's perception of the movie. In particular, high ratings often lead to high and long-lasting popularity.

% Given the inherent fluctuations in item popularity and the variable release timings of items, it is apparent that popular items change across different periods.
% Inspired by the herd mentality among users, we believe that the recent popularity of items plays a vital role in users' decision-making. 
% This was substantiated by \citet{Yitong2020_pop}, who conducted a comparative analysis of performances on the \textit{MovieLens} dataset.
% of the \textit{MostPop}, \textit{RecentPop}, and \textit{DecayPop} models. The \textit{MostPop} model recommends items with the highest global popularity, while \textit{RecentPop} recommends the most popular movies from the past month. The \textit{DecayPop} model, on the other hand, accounts for the weighted sum of an item's popularity over the past six months.
% Explain?
% Their findings demonstrated that recommending the most popular movies from the past month significantly outperformed recommending items with the highest global popularity, thereby underscoring the importance of recent item popularity in enhancing recommendation accuracy.
In this work, we introduce a straightforward model without complex network architectures, named \longname{} (\shortname{}). \shortname{} makes non-personalized recommendations by selecting the item predicted to have the highest popularity. \shortname{} relies simply on item features, including the popularity history and side information. Given the observed pattern of items experiencing boom and bust over time in Figure~\ref{fig:dataset_statistic}, we incorporate the current time as well as the item release time into \shortname{}. Besides, observing the periodic fluctuations in popularity experienced by different item genres in Figure \ref{fig:data_genre}, \shortname{} captures these periodic shifts to refine the predictive capability.

We perform comprehensive experiments on three real-world datasets to demonstrate the effectiveness of \shortname{}. Remarkably, the simplistic non-personalized \shortname{} performs on par or even better than the state-of-the-art sophisticated recommendation systems.
% It is worth noting that we use a fixed global time-point to split the dataset into training, validation, and testing sets. This splitting effectively prevents information leakage and more closely resembles real-world scenarios compared to commonly used strategies such as the Leave-One-Last or Temporal-User-Split \cite{Campos2022_split,Zaiqiao2020_split}.
Given that our proposed \shortname{} focuses on capturing item popularity to make recommendations, it differs from existing methods that target the capture of users' preferences. Therefore, \shortname{} can serve as a complementary component to enhance existing recommendation systems. We incorporated \shortname{} into existing personalized recommendation models and found that \shortname{} significantly enhances the performances of all baselines, including traditional recommendation methods, and the state-of-the-art non-sequential and sequential methods.
% mention Ablation study?

With this paper, we make the following contributions:
\begin{itemize}
    \item Consistent with \citet{Yitong2020_pop}, we found that recommending recently popular items performs better than recommending globally popular items to a large margin.  In particular, on \textit{Douban Movie}, it surpasses all recommendation baselines in terms of all metrics in the top 10 recommendations, further emphasizing the importance of recent item popularity.
    \item Observing the evolving item popularity over time and the herd mentality of users when making decisions, we propose to model item popularity trends over time. To the best of our knowledge, this is the first work explicitly predicting item popularity in recommendation systems. 
    \item Extensive experiments demonstrate the effectiveness of \shortname{}, which approximately doubles the NDCG@10 of the best state-of-the-art baselines on \textit{Douban Movies}. The simplicity of \shortname{} makes it a valuable baseline for future research, as well as a practical solution for industrial applications. 
    \item Further, when we integrate \shortname{} with existing recommendation models, the performance of this combined approach surpasses that of any individual model. Ablation studies show small overlaps in the recommendation lists generated by \shortname{} and the best baseline, ICLRec.
    These findings underscore the potential of \shortname{} as a powerful complement to existing recommendation systems. 
    % \item abalation study shows...
\end{itemize}

\section{Related Works}
\subsection{Sequential Recommendation}
Traditional recommendation methods often assign equal importance to all historical user-item interactions, overlooking the reality that user preferences and the appeal of items can change over time~\cite{he2016fusing,rendle2010factorizing, he2017translation}. Furthermore, recognizing sequential dependencies in user behaviors, such as purchasing car insurance after buying a car, can enrich the system's understanding of the user's actions. Therefore, sequential recommendation systems are introduced to capture the evolution of user preferences, which places a greater emphasis on recent interactions~\cite{ ma2019hierarchical, tang2018personalized, li2019translation, liu2018stamp}. 
% By accounting for the sequence of user behaviors, we can enhance the accuracy of recommendations.

Early sequential recommendation methods model the sequential patterns with Markov Chain (MC)~\cite{rendle2010factorizing, he2016fusing, he2016vista} or translation-based models~\cite{he2017translation, li2019translation}. \citet{rendle2010factorizing} combined the first-order MC and matrix factorization to model the sequential information and make predictions, achieving admirable results. \citet{he2016fusing} further introduced high-order MC to extract more complicated information to make personalized recommendations. \citet{he2017translation} proposed a Translation-based model, TransRec, which focuses on user-item-item third-order relationship.

Later, deep neural network approaches have been integrated into sequential recommendation systems. 
It is intuitive to utilize recurrent neural networks (RNNs) due to their capability to effectively process sequential inputs~\cite{hidasi2015session, hidasi2016parallel, donkers2017sequential, quadrana2017personalizing, ma2019hierarchical, yu2016dynamic}. RNN-based sequential recommendation systems usually leverage long-short-term-memory (LSTM) or gated recurrent units (GRU) to capture sequential dependencies~\cite{donkers2017sequential, zhu2017next, hidasi2015session, zheng2019gated}.
However, these RNN-based models heavily depend on interaction sequences and are tailored to model point-wise dependencies, potentially overlooking collective dependencies~\cite{kang2018self, sun2019bert4rec}.
Additionally, convolution neural networks (CNN) are also applied in sequential recommendation systems~\cite{tang2018personalized, jiang2022adamct}. These systems first regard sequential interaction as a matrix and subsequently treat this matrix as an "image" in both temporal and latent spaces~\cite{tang2018personalized}.
In recent years, graph neural networks (GNN) have emerged as a leading approach in sequential recommendation systems~\cite{wu2019session, zhang2020personalized, xu2019graph} and the attention mechanism has demonstrated significant promise in the sequential recommendation~\cite{li2017neural,liu2018stamp,kang2018self,fan2022sequential}.
% \citet{wu2019session} utilized a gated graph network to handle the neighborhood information in the user-item relation graph. To capture the user’s long-term performance, \citet{zhang2020personalized} introduced a Dot-Attention mechanism to explicitly model the impact of historical data at the current timestamp. 
For example, to model both global and local information on the graph, \citet{xu2019graph} dynamically constructed a graph with a self-attention mechanism for session sequences.

\textbf{Summary:} Sequential recommendation systems are designed to capture the change in users' preferences by assigning varying levels of importance to historical user-item interactions. However, to the best of our knowledge, all the existing sequential recommendation approaches fail to model the fluctuations in item popularity over time, which are crucial for influencing users' decisions.

\subsection{Item Popularity in Recommender Systems}

It is intuitive to make recommendations based on the items' popularity. The non-personalized strategy, which consistently recommends the most popular items according to the whole interaction history, has often been employed as a benchmark in assessing recommendation systems~\cite{Yitong2020_pop, tang2018personalized, kang2018self, ncf2017}. Also, item popularity has been discussed extensively in relation to recommendation systems.
% Additionally, there has also been a lot of discussion about the impact of item popularity on recommendation systems.

First, the so-called "long-tail" phenomenon presents a significant challenge. In this scenario, a small fraction of items gain immense popularity and attract a large user base, while a majority of items are consumed by very few users~\cite{abdollahpouri2019unfairness, Himan2020_bias}, which may lead the recommendation system to over-recommend popular items. Various methods have been proposed to address such problems, including regularization models~\cite{abdollahpouri2017controlling, Zhihong2020_bias, kamishima2014correcting, Ziwei2021_bias}, Causal-based models~\cite{wang2021deconfounded, wei2021model, zhang2021causal, zheng2021disentangling}, and adversarial models~\cite{Krishnan2018_bias, arduini2020adversarial}. 
Regularization models directly regulate the model predictions according to item popularity~\cite{Ziwei2021_bias, kamishima2014correcting, abdollahpouri2017controlling} or placing more emphasis on unpopular items~\cite{Zhihong2020_bias}. On the other hand, causal-based methods apply counterfactual intervention over the Causal Graph~\cite{peters2017elements} to mitigate the bias. Lastly, adversarial models try to strike a balance between recommending less popular items and using existing knowledge to maintain recommendation accuracy~\cite{Krishnan2018_bias, arduini2020adversarial}.
However, removing popularity bias directly usually negatively impacts the accuracy of the recommendations. As a result, recent studies have been focusing on reducing popularity bias while maintaining the models' performances~\cite{zhao2022popularity, xv2022neutralizing, yang2023debiased}.

Moreover, researchers have studied the effects of different methods of calculating item popularity on recommendation performance. 
\citet{Yitong2020_pop} compared the perofmances of the \textit{MostPop}, \textit{RecentPop}, and \textit{DecayPop} models. The \textit{MostPop} model recommends items with the highest global popularity, while \textit{RecentPop} recommends the most popular movies from the past month. The \textit{DecayPop} model, on the other hand, accounts for the weighted sum of an item's popularity over the past six months. The results showed that both \textit{RecentPop} and \textit{DecayPop} outperformed the traditionally used \textit{MostPop} model, suggesting that recent item popularity influences user choices more than overall item popularity throughout the entire interaction history. In another study, \citet{anelli2019local} proposed \textit{TimePop} to track item popularity within a user's specific network and make recommendations based on this personalized item popularity.

\textbf{Summary:}  Despite a wealth of research on item popularity in recommendation systems, there's a gap when it comes to explicitly predicting an item's future popularity trend. Furthermore, current research focuses on the interactions between users and items, often underestimating the influence of item popularity trends on the recommendation system.

\section{Popularity-Aware Recommender}
% In this section, we will first introduce the popularity forecasting task, then present the details of the proposed \shortname{}. 
\subsection{Problem Definition}
\begin{figure*}
	\centering
	\includegraphics[width=2\columnwidth]{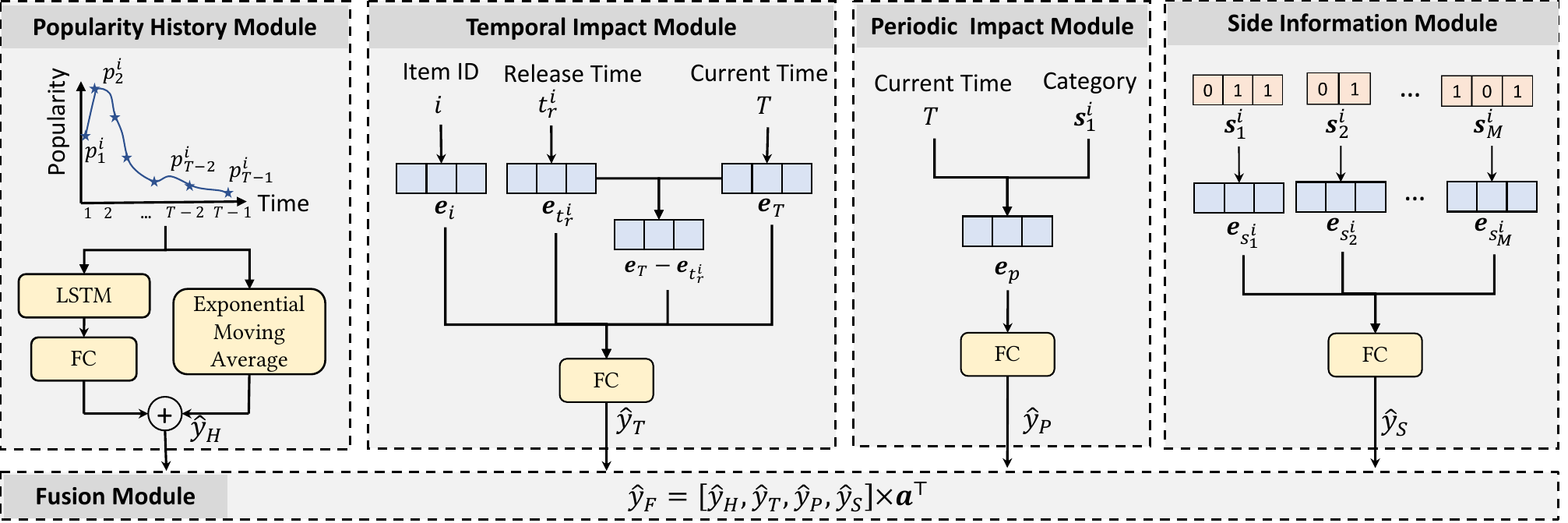}
	\caption{The proposed architecture of \shortname{}. The model consists of four modules modeling popularity history, temporal impact, periodic impact, and side information, respectively. Finally, an attention layer is leveraged to fuse the outputs of four modules. }
	\label{fig:arc}
\end{figure*}
We first present essential notations. We use $i\in \mathcal{I}$ to denote the item and $t\in \mathcal{T}$ to denote the time. 
Each item $i$ is associated with multiple features, such as the release time $t^i_r$ and $M$ other side information $\mathcal{S}^i=\{s^i_{1},\cdots, s^i_{M}\}$. Note that $\mathbf{s}^i_1\in \mathbb{R}^C$ refers to the item category, whereas $C$ refers to the number of categories. If item $i$ is associated with category $j$, $\mathbf{s}^i_1[j] = 1$, otherwise, $\mathbf{s}^i_1[j] = 0$.
Also, the time $t$ is not a single timestamp but a period of time, so item $i$ may have multiple interactions at time $t$. 
We denote the set of users who have interacted with item $i$ at time $t$ as $\mathcal{U}^i_t=\{u^i_{1},\cdots, u^i_{|\mathcal{U}^i_t|}\}$, where $|\cdot|$ denotes the set size. The popularity of $i$ is then defined as $p^i_t=|{\mathcal{U}^i_t}|$, where $t \geq t^i_r$.

Given the item release time $t^i_r\in \mathcal{T}$, the popularity history $\mathbf{p}^i = [p^i_{t^i_r},\cdots,p^i_{T-2},p^i_{T-1}]$, and other side information $\mathcal{S}^i$ of item $i$, the goal is to predict the item popularity $p^i_{T}$ at time $T$. Then the top $N$ items with the highest predicted popularity will be recommended to all users without distinction.

\subsection{Model Architecture}

As shown in Figure~\ref{fig:arc}, \shortname{} consists of four concise modules, each designed to predict the impending item popularity from distinct facets of item attributes. Finally, a Fusion Module combines the four predictions using an attention layer.

\subsubsection{Popularity History Module}

The past popularity of an item can usually provide a reasonable approximation of its current popularity.
We believe that an item's popularity generally follows specific trends based on its current popularity and rarely experiences sudden, drastic shifts over time.
Therefore, as shown in Figure~\ref{fig:arc}, we incorporated two simple but effective components into \shortname{}, which are designed to assess the item's latest popularity status and predict the impending popularity trend, respectively.
% We use the latest $K$ popularity for item $i$ at time $T$, represented as $\mathbf{p}^i =[p^i_{T-K},\cdots,p^i_{T-2},p^i_{T-1}]$.

First, given the presumption that recent popularity carries more significance than older popularity, we utilize the exponential moving average (EMA) \cite{klinker2011exponential} method, which assigns a higher weight to more recent data points. We use $\text{EMA}_{t}^i$ to denote the popularity estimator for item $i$ at time $t$:
% Therefore, $\text{EMA}_{t}^i$ can be defined as follows:
\begin{equation}
\text{EMA}_{t}^i = \left\{
    \begin{array}{ccl}
        0 & & \text{if} \ t < t^i_r \\
        p^i_{t^i_r} & & \text{if} \ t = t^i_r, \\
        \alpha p^i_{t} + (1-\alpha) \text{EMA}_{t-1}^i & & else
    \end{array}
\right.
\end{equation}
where $\alpha \in [0,1]$. A higher value of $\alpha$ refers to a greater emphasis on the recent popularity of the item. The latest popularity status at current time $T$ is then defined as $\hat{y}_{status} = \text{EMA}_{T-1}$.

Then, in order to predict whether an item will gain increased attention or decline in popularity, we utilize the Long Short-Term Memory (LSTM) method~\cite{sak_long_2014}. 
The distinctive use of a cell state and gating mechanisms in LSTMs allows them to selectively remember or forget information across various time intervals, rendering them particularly suitable for tasks involving long sequences.
More specifically, given the popularity history $\mathbf{p}^i = [p^i_{t^i_r},\cdots,p^i_{T-2},p^i_{T-1}]$, the following operations are carried out:
\begin{equation}
    \begin{split}
        I_t &= \sigma(\textbf{W}_{I}[p_{t}^i, H_{t-1}]+\textbf{b}_{I})\\
        F_t &= \sigma(\textbf{W}_{F}[p_{t}^i, H_{t-1}]+\textbf{b}_{F})\\
        G_t &= \text{tanh}(\textbf{W}_{G}[p_{t}^i, H_{t-1}]+\textbf{b}_{G})\\
        O_t &= \sigma(\textbf{W}_{O}[p_{t}^i, H_{t-1}]+\textbf{b}_{O})\\
        C_t &= F_t \odot C_{t-1} + I_t \odot G_t\\
        H_t &= O_t \odot \text{tanh}(C_t),
    \end{split}
\end{equation}
where $H_t$, and $C_t$ represent the hidden state and cell state of the popularity at time $t$. 
$I_t$, $F_t$, $G_t$, and $O_t$ denote the input, forget, cell, and output gates, respectively.
% $\textbf{W}_{I}$, $\textbf{W}_{F}$, $\textbf{W}_{G}$, $\textbf{W}_{O}$ , $\textbf{b}_{I}$, $\textbf{b}_{F}$, $\textbf{b}_{G}$, $\textbf{b}_{O}$ denote the weight matrix and bias vector for each gate, respectively. 
$\sigma$ denotes a sigmoid layer that maps the values between 0 and 1, with 1 meaning retaining the whole information and 0 signifying discarding it entirely. The operation $\odot$ denotes the Hadamard product.
The final hidden state, $H_{T-1}$, is then fed into a fully connected layer to predict the popularity trend:
\begin{equation}
    \hat{y}_{trend} = \boldsymbol{w}_H^\top H_{T-1}+\boldsymbol{b}_H.
\end{equation}
In this case, a positive $\hat{y}_{trend}$ suggests an increase in popularity, while a negative value indicates a decrease.

Finally, we combine the two estimations using:
\begin{equation}
    \hat{y}_H = \hat{y}_{status} + \hat{y}_{trend}.
    \label{equa:yih_and_tau}
\end{equation}

\subsubsection{Temporal Impact Module}

As illustrated in Figure \ref{fig:data_all}, we notice that peak attention typically occurs within the first two months after the item's release. Generally, as the temporal distance from the item's release increases, the item tends to diminish in popularity. The duration for which an item can maintain its popularity is significantly tied to the item itself. 
In this module, \shortname{} captures the influence of temporal factors on the item's popularity.

First, the current time $T$ and item release time $t_r^i$ are embedded, yielding $\mathbf{e}_T\in\mathbb{R}^d$ and $\mathbf{e}_{t_r^i}\in\mathbb{R}^d$, respectively. $d$ refers to the embedding size. Note that both times share the same embedding space. 
Given the significant role of temporal distance in popularity prediction, we define $\mathbf{e}_{dis} = \mathbf{e}_T - \mathbf{e}_{t_r^i}$. Then we concatenate $\mathbf{e}_T$, $\mathbf{e}_{t_r^i}$, $\mathbf{e}_{dis}$, along with the item embedding $\mathbf{e}_i\in\mathbb{R}^d$. Finally, the concatenation is fed into a fully connected layer to predict the item popularity:
\begin{equation}
    \begin{split}
        % \boldsymbol{e}^T_{i,t} &= \text{Concat}(\boldsymbol{e}^T_t, \boldsymbol{e}^{T}_{r_i}, \boldsymbol{e}^T_t - \boldsymbol{e}^{T}_{r_i}, \boldsymbol{e}^I_i ),\\
        \hat{y}_{T} &= \text{ReLU}(\boldsymbol{w}_T^\top [\mathbf{e}_T, \mathbf{e}_{t_r^i}, \mathbf{e}_{dis}, \mathbf{e}_i] + \boldsymbol{b}_T).
    \end{split}
\end{equation}

\subsubsection{Periodic Impact Module}

Besides the temporal evolution of item popularity, we also observe periodic fluctuations in the popularity of different categories of items over time. As shown in Figure \ref{fig:data_genre}, movies from \textit{Romance} and \textit{Animation} undergo periods of surges and declines in different months. This phenomenon is not exclusive to movies and can be seen across various items. For instance, T-shirts see increased popularity during summer, whereas sweaters gain popularity during winter.

To capture this periodic effect, we construct an embedding matrix $\mathbf{E}\in\mathbb{R}^{(\omega C)\times d}$, where $\omega$ represents period time. For example, if each $t$ refers to one month and $\omega=12$, it suggests that categories of an item follow similar popularity trends annually. 
Given the current time $T$, we first transform to a one-hot vector $\mathbf{v}_T\in \{0,1\}^\omega$. If time $T$ falls within the $j^{th}$ period, then $\mathbf{v}_T[j]=1$, otherwise, $\mathbf{v}_T[j]=0$.
Then we calculate the periodic embedding using the item categories $\mathbf{s}^i_1 \in \mathbb{R}^C$ and time $\mathbf{v}_T$:
\begin{equation}
    \begin{split}
        \mathbf{e}_p &= \left[\text{Flatten}\left(\left(\mathbf{s}^i_1\right)^\top \times \mathbf{v}_T\right)\right]\times \mathbf{E}.
        % \boldsymbol{e}^P_i &= \text{Embed}(m*|\mathcal{G}|+G_i),\\
        % \hat{y}^P_i &= \text{ReLU}(\boldsymbol{w}^\top \boldsymbol{e}^P_i + \boldsymbol{b}),
    \end{split}
\end{equation}
Finally, we feed the embedding into a fully connected layer, which is activated by ReLU:
\begin{equation}
        \hat{y}_P = \text{ReLU}(\boldsymbol{w}_P^\top \boldsymbol{e}_p + \boldsymbol{b}_P).
\end{equation}

\subsubsection{Side Information Module}
Incorporating side information substantially enhances the accuracy of item popularity predictions. First, item's side information is closely related to the peak popularity and duration of popularity. Taking movies as an example, a movie produced by more famous directors or actors usually tends to attract a larger audience and stay popular for a longer period of time. Furthermore, side information can be particularly useful when dealing with cold-start items that have little or even no popularity history. For each kind of side information $\mathbf{s}_j \in \{0,1\}^{q_j}$, where $j\leq M$ and $q_j$ is the number of attributes for $j$, we feed them into an embedding layer:
\begin{equation}
        \mathbf{e}_{s_j^i} = \mathbf{s}^i_j \times \mathbf{E}_j,
\end{equation}
where $\mathbf{E}_j\in \mathbb{R}^{q_j\times d}$. Finally, we concatenate the $M$ kinds of side information and utilize a fully connected later to predict how the side information influences the item popularity:
\begin{equation}
        \hat{y}_S = \text{ReLU}(\boldsymbol{w}_S^\top [\mathbf{e}_{s_1^i}, \mathbf{e}_{s_2^i}, \cdots, \mathbf{e}_{s_M^i}] + \boldsymbol{b}_S).
\end{equation}

\subsubsection{Module fusion}
In this module, we combine the four predictions with a simple 4-dimensional attention vector $\boldsymbol{a}\in\mathbb{R}^4$:
\begin{equation}
    \hat{y}_F = [\hat{y}_H,\hat{y}_T,\hat{y}_P,\hat{y}_S] \times \boldsymbol{a}^\top
    % \hat{y}_{i,T} = a_1\hat{y}_H + a_2\hat{y}_T + a_3\hat{y}_P + a_4\hat{y}_S,
    \label{equa:fuse}
\end{equation}
subject to the condition $\sum \mathbf{a} = 1$. Through the attention layer, \shortname{} can incorporate all model factors, including item popularity history, temporal effects, periodic effects, and side information. Furthermore, the attention layer enhances interpretability by demonstrating how the four modules contribute to the final prediction.

\subsection{Training}
Since the predicted output from each module is expected to closely align with the actual item popularity, we employ the mean square error (MSE) loss from each module to train the model as follows:
% To train the model, we combine the Mean Square Error (MSE) loss of each module prediction as follows:
\begin{equation}
    \begin{split}
        \mathcal{L} = \sum_{j\in\{H,T,P,S,F\}} \mathbb{E}_{i\in\mathcal{I},t<T}\left[\left(p^i_t-\hat{y}_j\right)^2\right], \\ 
    \end{split}
    \label{equa:loss}
\end{equation}
where $p^i_t$ represents the actual popularity of item $i$ at time $t$, while $\hat{y}_j$ denotes the predicted popularity for item $i$ at time $t$, as generated by module $j$. We use Adam~\cite{kingma2014adam} for model optimization.

\section{Experimental Settings}
% In this section, we first introduce our post-process experiment setting that combines our popularity forecasting model with baseline model, then we present and analyse the experiment results.
\subsection{Datasets}

The proposed \shortname{} is evaluated on three real-world datasets: \textit{Douban Movies}, \textit{Video Games}, and \textit{Home and Kitchen}. The \textit{Douban Movies} dataset is crawled from the Douban website\footref{douban}, which is one of the largest Chinese social media sites that allow users to make comments on movies, books, music, etc. We crawl all movies that are released from 1$^{st}$ January 2011 to 31$^{st}$ December 2020 and their side information including categories, directors, and actors. Moreover, \textit{Douban Movies} consists of all user-item interactions where the user has commented on the item during the same time period. In summary, \textit{Douban Movies} comprises 33,635 users and 2,795 movies. 
Further evaluation is conducted using two public datasets from Amazon\cite{amazon}: \textit{Video Games} and \textit{Home and Kitchen}. More statistics are summarized in Table \ref{table:dataset_statistic}.

It is worth noting that we use a fixed global time-point to split the dataset into training, validation, and testing sets. This splitting effectively prevents information leakage and more closely resembles real-world scenarios compared to commonly used strategies such as the Leave-One-Last or Temporal-User-Split \cite{Campos2022_split,Zaiqiao2020_split}. 
In our experiments, we define each time $t$ as a 30-day period, with $t=1$ representing the first 30 days subsequent to the first user-item interaction in the whole dataset. The interactions from the final month are used for testing, those from the penultimate month are used for validation, and all remaining interactions are used for training.
% More specifically, let $t_{max}$ denote the time period of the last user-item interaction, we define the validation time-point and test time-point as $t_{valid} = t_{max}-1$, $t_{test} = t_{max}$. Any interactions occurring during $t_{test}$ are allocated for testing, while those occurring before $t_{valid}$ are used for training.
It's important to note that the time period in the test set may be less than one month, resulting in a relatively small number of interactions.

\begin{table}
    \centering
    \caption{Statistics of the datasets.}
    \label{table:dataset_statistic}
    \resizebox{\columnwidth}{!}{
    \begin{tabular}{ccccccc}
    \toprule
    \textbf{Dataset}  & \textbf{\#Users} & \textbf{\#Items} & \textbf{\#Train}  & \textbf{\#Validate}  & \textbf{\#Test} \\ 
    \midrule
    Douban Movies  & 33,635 & 2,795 & 329,380 & 1,992 & 560 \\ 
    Video Games  & 23,933 & 4,211 & 169,845 & 1,968 & 1,977\\
    Home and Kitchen  & 65,588 & 8,633 & 361,005 & 13,448 & 376\\
    \bottomrule
    \end{tabular}
    }
\end{table}
\subsection{Baselines}

\begin{table*}[t!]
    \centering
    \caption{Model performances of Top-10 recommendation. The best results among variants of Cutoff TopPop are marked with $^*$. The best results and the second-best results within each group are bold and underlined, respectively. Relative Imp-1. denotes to the improvement of \shortname{} over the best original baselines, Relative Imp-2 denotes the improvement of integrated model over \shortname{}.}
    \label{tab:baseline_5metric}
    \resizebox{\textwidth}{!}{ 
    \begin{tabular}{clccccccccccccccc}
    \toprule
\multicolumn{2}{c}{\multirow{2}{*}{Methods}}              & \multicolumn{5}{c}{Douban Movies}                                                                                                                           & \multicolumn{5}{c}{Video   Games}                                                                                                                        & \multicolumn{5}{c}{Home   and Kitchen}                                                                                                                 \\ \cmidrule{3-17}
\multicolumn{2}{c}{}                                      & Precision                   & Recall                      & HR                          & MRR                         & \multicolumn{1}{l}{NDCG}    & Precision                    & Recall                       & HR                           & MRR                          & \multicolumn{1}{l}{NDCG}     & Precision                    & Recall                       & HR                           & MRR                         & \multicolumn{1}{l}{NDCG}    \\ \midrule
\multirow{4}{*}{\begin{tabular}[c]{@{}c@{}}Cutoff \\ TopPop\end{tabular}} & 3 months                        & 0.0177$^*$                      & 0.1460$^*$                      & 0.1742$^*$                      & 0.0062$^*$                      & 0.0661$^*$                      & 0.0086                       & 0.0565                       & 0.0833                       & 0.0039                       & 0.0285$^*$                       & 0.0039$^*$                       & 0.0370$^*$                       & 0.0391  $^*$                     & 0.0036$^*$                      & 0.0335$^*$                      \\
                               & 6 months                       & 0.0118                      & 0.0822                      & 0.1124                      & 0.0045                      & 0.0387                      & 0.0092                       & 0.0567$^*$                       & 0.0877                       & 0.0041$^*$                       & 0.0283                       & 0.0004                       & 0.0043                       & 0.0043                       & 0.0002                      & 0.0027                      \\
                               & 12 months                      & 0.0115                      & 0.0796                      & 0.1096                      & 0.0032                      & 0.0316                      & 0.0094$^*$                       & 0.0545                       & 0.0892 $^*$                     & 0.0040                       & 0.0274                       & 0.0000                       & 0.0000                       & 0.0000                       & 0.0000                      & 0.0000                      \\
                               & ALL                      & 0.0042                      & 0.0306                      & 0.0421                      & 0.0012                      & 0.0119                      & 0.0069                       & 0.0391                       & 0.0643                       & 0.0036                       & 0.0218                       & 0.0000                       & 0.0000                       & 0.0000                       & 0.0000                      & 0.0000                      \\\midrule
\multirow{11}{*}{Original}    
                               & UserKNN                  & 0.0048                      & 0.0345                      & 0.0449                      & 0.0016                      & 0.0143                      & 0.0127                       & 0.0809                       & 0.1155                       & 0.0044                       & 0.0359                       & 0.0048                       & 0.0384                       & 0.0478                       & 0.0015                      & 0.0174                      \\ 
                               & ItemKNN                  & 0.0039                      & 0.0252                      & 0.0337                      & 0.0015                      & 0.0110                      & 0.0121                       & 0.0763                       & 0.1097                       & 0.0042                       & 0.0338                       & 0.0043                       & 0.0370                       & 0.0435                       & 0.0018                      & 0.0193                      \\
                               & SLIM BPR                 & 0.0051                      & 0.0360                      & 0.0506                      & 0.0018                      & 0.0159                      & 0.0117                       & 0.0706                       & 0.1067                       & 0.0044                       & 0.0326                       & 0.0043                       & 0.0391                       & 0.0435                       & 0.0014                      & 0.0177                      \\
                               & NCF                      & 0.0045                      & 0.0318                      & 0.0421                      & 0.0014                      & 0.0118                      & 0.0066                       & 0.0377                       & 0.0599                       & 0.0033                       & 0.0198                       & 0.0026                       & 0.0217                       & 0.0261                       & 0.0007                      & 0.0094                      \\
                               & SHT                      & 0.0059	& 0.0364	& 0.0590	& 0.0028	& 0.0193                    & 0.0102                       & 0.0600                       & 0.0921                       & 0.0039                       & 0.0291                       & 0.0022                       & 0.0217                       & 0.0217                       & 0.0007                      & 0.0100                      \\
                               & Caser                    & 0.0048                      & 0.0339                      & 0.0478                      & 0.0010                      & 0.0116                      & 0.0089                       & 0.0620                       & 0.0863                       & 0.0026                       & 0.0248                       & 0.0043                       & 0.0362                       & 0.0435                       & 0.0010                      & 0.0147                      \\
                               & SASRec                   & 0.0098                      & 0.0691                      & 0.0955                      & 0.0037                      & 0.0336                      & 0.0089                       & 0.0497                       & 0.0863                       & 0.0043                       & 0.0269                       & 0.0035                       & 0.0326                       & 0.0348                       & 0.0017                      & 0.0195                      \\
                               & HGN                      & 0.0104                      & 0.0713                      & 0.0983                      & 0.0041                      & 0.0345                      & {\ul 0.0151}                 & {\ul 0.1005}                 & 0.1389                       & {\ul 0.0053}                 & \textbf{0.0465}                 & \textbf{0.0074}              & {\ul 0.0667}                 & 0.0696                       & 0.0019                      & 0.0277                      \\
                               & STOSA                    & 0.0124                      & 0.0870                      & 0.1208                      & {\ul 0.0062}                & {\ul 0.0484}                & 0.0148	& 0.0941	& {\ul 0.1404}	&  {\ul 0.0053}	& 0.0425 & \textbf{0.0074}              & 0.0623                       & \textbf{0.0739}              & {\ul 0.0031}                & {\ul 0.0323}                \\
                               & ICLRec                   & {\ul 0.0135}                & {\ul 0.0961}                & {\ul 0.1236}                & 0.0054                      & 0.0456                      & \textbf{0.0178}              & \textbf{0.1125}              & \textbf{0.1667}              & \textbf{0.0056}              & {\ul 0.0430}                       & {\ul 0.0070}                 & 0.0623                       & 0.0696                       & 0.0024                      & 0.0279                      \\
                               & \multicolumn{1}{l}{\shortname{}} & \textbf{0.0208}             & \textbf{0.1695}             & \textbf{0.1994}             & \textbf{0.0092}             & \textbf{0.0955}             & 0.0104                       & 0.0647                       & 0.0950                       & 0.0042                       & 0.0300                       & \textbf{0.0074}              & \textbf{0.0674}              & {\ul 0.0696}                 & \textbf{0.0054}             & \textbf{0.0527}             \\ \midrule
\multicolumn{2}{c}{Relative Imp-1.}                        & 54.17\%         & 76.46\%         & 61.36\%         & 47.64\%         & 97.24\%                  & -41.80\%        & -42.51\%        & -42.98\%        & -25.42\%        & -35.59\%                 & 0.00\%          & 1.09\%          & -5.88\%         & 74.56\%         & 62.80\%                  \\ \bottomrule
\multirow{10}{*}{+\shortname{}}       
                               & UserKNN                  & 0.0208                      & 0.1688                      & 0.2023                      & {\ul 0.0092}                & {\ul 0.0946}                & 0.0154                       & 0.0922                       & 0.1418                       & 0.0056                       & 0.0439                       & 0.0091                       & 0.0804                       & 0.0826                       & 0.0054                      & 0.0540                      \\  & ItemKNN                  & {\ul 0.0225}                & 0.1716                      & 0.2079                      & 0.0084                      & 0.0833                      & 0.0155                       & 0.0994                       & 0.1433                       & 0.0052                       & 0.0420                       & 0.0091                       & 0.0804                       & 0.0826                       & 0.0056                      & 0.0558                      \\
                               & SLIM BPR                 & 0.0211                      & 0.1716                      & 0.2051                      & 0.0086                      & 0.0927                      & 0.0149                       & 0.0945                       & 0.1374                       & 0.0055                       & 0.0425                       & 0.0087                       & 0.0783                       & 0.0826                       & 0.0053                      & 0.0529                      \\
                               & NCF                      & 0.0219                      & 0.1702                      & 0.2079                      & 0.0085                      & 0.0859                      & 0.0096                       & 0.0595                       & 0.0921                       & 0.0041                       & 0.0284                       & 0.0083                       & 0.0761                       & 0.0783                       & 0.0044                      & 0.0474                      \\
                               & SHT                      & 0.0216	& 0.1735	& 0.2107	& 0.0085	& 0.0798           & 0.0140                       & 0.0814                       & 0.1272                       & 0.0050                       & 0.0362                       & 0.0083                       & 0.0761                       & 0.0783                       & 0.0041                      & 0.0451                      \\
                               & Caser                    & 0.0213                      & 0.1721                      & 0.2135                      & 0.0073                      & 0.0785                      & 0.0116                       & 0.0703                       & 0.1067                       & 0.0042                       & 0.0309                       & 0.0117                       & 0.1109                       & 0.1130                       & 0.0053                      & 0.0613                      \\
                               & SASRec                   & 0.0225                      & {\ul 0.1751}                & {\ul 0.2163}                & 0.0074                      & 0.0772                      & 0.0111                       & 0.0659                       & 0.1038                       & 0.0045                       & 0.0308                       & 0.0074                       & 0.0674                       & 0.0696                       & {\ul 0.0057}                & 0.0543                      \\
                               & HGN                      & \textbf{0.0236}             & \textbf{0.1819}             & \textbf{0.2247}             & 0.0087                      & 0.0858                      & {\ul 0.0181}                 & {\ul 0.1151}                 & {\ul 0.1725}                 & {\ul 0.0059}                       & {\ul 0.0493}                       & {\ul 0.0135}                 & {\ul 0.1239}                 & {\ul 0.1261}                 & {\ul 0.0057}                & {\ul 0.0657}                \\
                               & STOSA                    & 0.0222                      & 0.1730                      & 0.2135                      & 0.0080                      & 0.0807                      & 0.0151	& 0.1001	& 0.1433	& 0.0055	& 0.0453                 & \textbf{0.0148}              & \textbf{0.1348}              & \textbf{0.1391}              & \textbf{0.0069}             & \textbf{0.0776}             \\ 
                               & ICLRec                   & 0.0208                      & 0.1695                      & 0.1994                      & \textbf{0.0094}             & \textbf{0.0970}             & \textbf{0.0194}              & \textbf{0.1205}              & \textbf{0.1813}              & \textbf{0.0068}              & \textbf{0.0508}              & 0.0122                       & 0.1080                       & 0.1130                       & 0.0056                      & 0.0618                      \\\midrule
\multicolumn{2}{c}{Relative Imp-2.}                          & 13.51\%         & 7.32\%          & 12.68\%         & 2.52\%          & 1.54\%                   & 87.32\%         & 86.32\%         & 90.77\%         & 63.21\%         & 69.52\%                  & 100.01\%        & 100.00\%        & 100.00\%        & 27.46\%         & 47.38\%  \\ \bottomrule
\end{tabular}
    
    }
\end{table*}

We select the following methods for evaluation, including traditional methods, non-sequential recommendation methods, and sequential recommendation methods. We use the published codes for implementing baseline methods.
% Traditional recommendation baselines (i.e., UserKNN \cite{sarwar2001item}, ItemKNN \cite{wang2006unifying}, SLIM \cite{slim})\footnote{https://github.com/MaurizioFD/RecSys2019\_DeepLearning\_Evaluation} are implemented following the work of \cite{recsys19}.

\begin{itemize}
    \item \textbf{Cutoff TopPop}, which recommends the items that are most popular during a specific time period. \textit{Cutoff TopPop -- ALL} is one of the most widely used recommendation baselines that always recommend the most popular item based on the entire history of user-item interactions. Inspired by \citet{Yitong2020_pop}, we also incorporate variations of TopPop that calculate popularity using recent interaction data. In our experiments, we consider recent interactions in the last 3 months, 6 months, and 12 months.
    \item \textbf{UserKNN}\footnote{\href{https://github.com/MaurizioFD/RecSys2019_DeepLearning_Evaluation}{https://github.com/MaurizioFD/RecSys2019\_DeepLearning\_Evaluation}\label{baseline:trad}} \cite{sarwar2001item,recsys19}, a traditional collaborative filtering method based on the k-nearest neighbors algorithm. Using the validation set, we select the best from five distance functions for user-user similarity, including cosine, Jaccard, Dice, Tversky, and asymmetric distances.
    \item \textbf{ItemKNN}\footref{baseline:trad} \cite{wang2006unifying}, a neighborhood-based method using collaborative item-item similarity. We tune the selection of the distance functions in the same manner as UserKNN. 
    \item \textbf{SLIM BPR}\footref{baseline:trad} \cite{ning2011slim}, which leverages a sparse coefficient matrix to predict user behaviors and is optimized with Bayesian Personalized Ranking (BPR) loss.
    \item \textbf{NCF}\footnote{\href{https://github.com/hexiangnan/neural_collaborative_filtering}{https://github.com/hexiangnan/neural\_collaborative\_filtering}} \cite{ncf2017}, which is a non-sequential baseline that combines a generalized matrix factorization module and a multi-layer perceptron.
    \item \textbf{SHT}\footnote{\href{https://github.com/akaxlh/SHT}{https://github.com/akaxlh/SHT}} \cite{xia2022self}, which is a non-sequential baseline that replaces the dot-product in conventional matrix factorization with a hypergraph transformer network and conducts data augmentation with a generative self-supervised learning component.
    \item \textbf{Caser}\footnote{\href{https://github.com/graytowne/caser}{https://github.com/graytowne/caser}} \cite{tang2018personalized}, which is a CNN-based method that applies horizontal and vertical convolutions for sequential recommendation.
    \item \textbf{SASRec}\footnote{\href{https://github.com/kang205/SASRec}{https://github.com/kang205/SASRec}} \cite{kang2018self}, which is a self-attention-based sequential model which utilizes an attention mechanism. 
    \item \textbf{HGN}\footnote{\href{https://github.com/allenjack/HGN}{https://github.com/allenjack/HGN}} \cite{ma2019hierarchical}, which captures both user intents and item-item relations from item sequences with a hierarchical gating network. 
    \item \textbf{STOSA}\footnote{\href{https://github.com/zfan20/STOSA.}{https://github.com/zfan20/STOSA}} \cite{fan2022sequential}, which embeds each item as a stochastic Gaussian distribution, and forecasts the next item for sequential recommendation with a self-attention mechanism.
    \item \textbf{ICLRec}\footnote{\href{Code is available at https://github.com/salesforce/ICLRec}{https://github.com/salesforce/ICLRec}} \cite{chen2022intent}, which learns users’ preferences from unlabeled user historical interactions and is optimized through contrastive self-supervised learning.
    
\end{itemize}

\subsection{Evaluation Metrics}
To evaluate the performance of Top-N recommendations, we use the hit ratio (HR), precision, recall, mean reciprocal rank (MRR), and normalized discounted cumulative gain (NDCG) for N = 1,3,5,7,10. The HR, precision, and recall metrics measure whether and how many of the target item appears in the top-N list, whereas the MRR and NDCG consider the ranking position of target items within the list.

% We adopt tow common Top-N metrics in our experiment, \textit{Hit Rate@k} (HR@k) and \textit{NDCG@k} (NDCG@k) \cite{kang2018self}, $k \in \{5, 10\}$. HR@k counts the fraction of times that the ground-truth item occurs in the top k items. NDCG@k is the normalized discounted cumulative gain at k, which considers the position of correctly recommended items by assigning larger weights on higher positions. 

\subsection{Method Integration}
\label{sec:post-processing-method}
To demonstrate the efficacy of the \shortname{} as a complementary component to existing recommendation methods, we propose to incorporate the item popularity score, as predicted by our model, with the estimated user preferences for items in existing recommendation methods.
More specifically, if $\hat{y}_F$ represents the predicted item popularity for item $i$ at time $T$, $s(u,i)$ represents the predicted preference score for user $u$ regarding item $i$, with a higher $s(u,i)$ indicating a greater likelihood for the user to select the item. Then the updated ranking score at a specific time $T$ can be formulated as follows:
\begin{equation}
    s_{new}(u, i, T) = \beta s(u,i) + (1-\beta) \hat{y}_F,
    \label{equa:integration}
\end{equation}
where $\beta$ serves as a hyperparameter that regulates the balance between the influence of item popularity and user personalization in user decisions. 
A lower value of $\beta$ indicates that user choices are more influenced by the current item popularity, while a higher $\beta$ emphasizes more on the importance of personalized recommendations.

\subsection{Implementation Details}
% \hl{For Jing Jiazheng: pls modify this part}
In our experiments, we select learning rate from $[0.0001, 0.1]$ and batch size from $\{64, 128, 256\}$. We also apply L2 regularization when computing the loss function in Equation \ref{equa:loss} with the weight decay being $0.0001$. The embedding size $d$ is set to 64.\footnote{Our code is available at \href{https://github.com/JingXiaoyi/PARE}{https://github.com/JingXiaoyi/PARE}.}
% We implement our \shortname{} with PyTorch and train the model with the Loss function (\ref{equa:loss}). We select learning rate from $\{1e-1, 1e-2, 1e-3, 1e-4\}$ and  $\lambda=1e-4$. Besides, we select training batch size from $\{64, 128, 256\}$ and set the embedding size to 64. For initialization, we set $\boldsymbol{a} = [0.25, 0.25, 0.25, 0.25]$ and randomly initialized all the hidden layers to Normal distribution with $mean=0$, and $std=0.1$.

\section{Results and Discussion}
We first compare the performances between different recommendation models as well as their variants integrated with \shortname{}.
We also carry out a comprehensive ablation study to assess the effectiveness of each module within our proposed \shortname{}, the accuracy of item popularity prediction, as well as the efficacy of our model when used as a complementary component alongside existing recommendation baselines.

\begin{table}[t!]
    \centering
    \caption{Model performances with different Top-N value of hit ratio (HR) on \textit{Douban Movies}. The best results among variants of Cutoff TopPop are marked with $^*$. The best results and the second-best results within each group are bold and underlined, respectively. Relative Imp-1. denotes to the improvement of \shortname{} over the best original baselines, Relative Imp-2 denotes the improvement of integrated model over \shortname{}.}
    \label{tab:baseline_douban}
    \resizebox{\columnwidth}{!}{ 
    \begin{tabular}{clcccc} \toprule
\multicolumn{2}{c}{Methods}                                                                       & @1               & @3               & @5               & @7               \\ \midrule
\multirow{4}{*}{\begin{tabular}[c]{@{}c@{}}Cutoff   \\      TopPop\end{tabular}} & 3 months  & 0.0337$^*$          & 0.0534          & 0.0646          & 0.1320$^*$          \\
                                                                                 & 6 months  & 0.0112          & 0.0674$^*$          & 0.0927$^*$          & 0.0955          \\
                                                                                 & 12 months & 0.0056          & 0.0506          & 0.0590          & 0.0843          \\
                                                                                 & ALL       & 0.0056          & 0.0112          & 0.0112          & 0.0309          \\ \midrule
\multirow{11}{*}{Original}                                                      
                                                                                 & UserKNN   & 0.0028          & 0.0197          & 0.0309          & 0.0393          \\
                                                                                 & ItemKNN   & 0.0084          & 0.0140          & 0.0253          & 0.0253          \\
                                                                                 & SLIM BPR  & 0.0112          & 0.0169          & 0.0197          & 0.0281          \\
                                                                                 & NCF       & 0.0056          & 0.0169          & 0.0281          & 0.0337          \\
                                                                                 & SHT       & 0.0169	& 0.0309	& 0.0393	& 0.0478         \\
                                                                                 & Caser     & 0.0000          & 0.0084          & 0.0253          & 0.0309          \\
                                                                                 & SASRec    & 0.0169          & 0.0449          & 0.0646          & 0.0646          \\
                                                                                 & HGN       & 0.0225          & 0.0421          & 0.0702          & 0.0787          \\
                                                                                 & STOSA     & {\ul 0.0393}    & {\ul 0.0702}    & {\ul 0.0983}    & 0.1039          \\
                                                                                 & ICLRec    & 0.0253          & {\ul 0.0702}    & 0.0899          & {\ul 0.1067}    \\
                                                                                 & \shortname{}      & \textbf{0.0534} & \textbf{0.0927} & \textbf{0.1461} & \textbf{0.1629} \\ \midrule
\multicolumn{2}{c}{Relative Imp-1.}                                                            & 35.71\%         & 32.00\%         & 48.57\%         & 52.63\%         \\ \midrule
\multirow{10}{*}{+ \shortname{}}                                                        & UserKNN   & 0.0309          & \textbf{0.1264}    & 0.1433          & 0.1714          \\
& ItemKNN   & 0.0337          & 0.1124          & {\ul 0.1629}          & 0.1770          \\
                                                                                 
                                                                                 & SLIM BPR  & 0.0225          & {\ul0.1208}          & 0.1517          & 0.1685          \\
                                                                                 & NCF       & \textbf{0.0562} & 0.1124          & 0.1461          & 0.1685          \\
                                                                                 & SHT       & 0.0309	& 0.1152	& 0.1545	& 0.1685          \\
                                                                                 & Caser     & 0.0281          & 0.0843          & 0.1433          & 0.1742          \\
                                                                                 & SASRec    & 0.0225          & 0.0871 & 0.1601          & 0.1798          \\
                                                                                 & HGN       & {\ul 0.0365}    & 0.1152          & \textbf{0.1685} & {\ul 0.1910}    \\
                                                                                 & STOSA     & 0.0253          & 0.1096          & 0.1573          & 0.1714          \\
                                                                                 & ICLRec    & 0.0337          & 0.1039          & {\ul 0.1629}    & \textbf{0.1938} \\ \midrule
\multicolumn{2}{c}{Relative Imp-2.}                                                            & 5.26\%          & 36.36\%         & 15.38\%         & 18.97\%         \\ \bottomrule
\end{tabular}

    }
\end{table}

\subsection{Recommendation Performances}
Due to space constraints, we present the model performance for the top 10 recommendations, as assessed by all evaluation metrics, across the three datasets in Table \ref{tab:baseline_5metric}. Besides, the hit ratio performance with varying top $N$ values on the \textit{Douban Movies} dataset is shown in Table \ref{tab:baseline_douban}. From these results, we make the following observations.

First, our findings align with those of \citet{Yitong2020_pop}, wherein recommending items based solely on recent popularity outperforms the commonly used \textit{Cutoff TopPop - ALL} baseline on all datasets in terms of all metrics. The latter always recommends the most popular item according to the entire interaction history. On the \textit{Douban Movies} dataset, the MRR@10 and NDCG@10 for \textit{Cutoff TopPop - 3 months} are more than four times greater than those for making recommendations considering all data.
Besides, we also find that the best variant of \textit{Cutoff TopPop} outperforms most recommendation baselines on \textit{Douban Movies}.
These results underline the significant influence of recent item popularity on user decision-making. Moreover, these findings inspire us to consider utilizing \textit{Cutoff TopPop} over a short time span, rather than the generally adopted \textit{Cutoff TopPop - ALL}, as a solid baseline for evaluating future recommendation models.

Second, the simple non-personalized \shortname{} performs at a similar level or even surpasses the more complex state-of-the-art recommendation methods. On \textit{Douban Movies}, \shortname{} outperforms the best baseline by $97.24\%$ and $76.46\%$ in NDCG@10 and Recall@10, respectively. However, we observed that \shortname{} performs relatively less effectively on the $\textit{Video Games}$ dataset, where the personalized sequential recommendation baseline \textit{ICLRec} exhibits the best performance. This could be attributed to the diverse tastes of users in video games, who tend to purchase based on their personal preferences rather than opting for the most popular choices. Considering these findings, our model could serve as a strong baseline for future recommendation evaluations and assist in analyzing the impact of both user preference and item popularity on user decision-making.

Third, when we integrate our proposed \shortname{} into existing recommendation methods using Equation \ref{equa:integration}, the combined model outperforms all corresponding original personalized recommendation baselines. 
We observe a relatively significant improvement over the original existing baselines on the \textit{Douban Movies} dataset and a more pronounced enhancement over \shortname{} on the other two datasets.
On \textit{Douban Movies}, the combined \textit{ItemKNN+\shortname{}} improves the original \textit{ItemKNN} by approximately 7 times in HR@3, however, it only enhances the original \shortname{} model by only about $21\%$. On \textit{Home and Kitchen}, we observe a doubling of performance when we incorporate existing recommendation baselines into our model in terms of precision, recall, and HR for the top 10 recommendations.

In summary, these experimental findings demonstrate the effectiveness of our proposed model both as a strong baseline for future research and as a potential complementary component capable of enhancing the performance of existing recommendation methods.

% \subsection{Ablation Study}

\subsection{Effectiveness of Each Module}
\begin{table}[]
    \centering
    \caption{Comparison of each module on \textit{Douban Movies}. H, T, P, and S denote the Popularity History Module, Temporal Impact Module, Side Information Model, and Periodic Impact Module, respectively. "Attention Weight" denotes the corresponding attention score in the Fusion Module.}
    \label{tab:module_compare}
    \begin{tabular}{lcccccc}\toprule
\multirow{2}{*}{Metric} & \multirow{2}{*}{HR@10} & \multicolumn{4}{c}{Attention Weight} &  \\ \cmidrule{3-7}
                        &                        & H       & T       & S       & P      &  \\ \midrule
H                       & 0.1798                 & 1.0000  & -  & -  & - &  \\
H+T                     & 0.1910                 & 0.6025  & 0.3975  & -  & - &  \\
H+T+S                   & 0.1938                 & 0.6070  & 0.2047  & 0.1882  & - &  \\
H+T+P                   & { 0.1966}           & 0.5980  & 0.1844  & -  & 0.2177 &  \\
H+T+S+P                 & 0.1994        & 0.6259  & 0.2573  & 0.0422  & 0.0746 & \\ \bottomrule
\end{tabular}
\end{table}

Comparing the variants of \textit{Cutoff TopPop} with \shortname{}, as seen in Table \ref{tab:baseline_5metric} and Table \ref{tab:baseline_douban}, we found that \shortname{} consistently outperforms \textit{Cutoff TopPop} across all metrics and datasets. This suggests that item popularity history is not the sole determinant of user decisions; additional item side information, such as categories and release times, are also key factors. 
To further explore the efficacy of each module, we compared the performance of recommendations when different module combinations were applied on \textit{Douban Movies}.

We illustrate the comparison results in Table \ref{tab:module_compare}, where H, T, S, and P denote the Popularity History Module, Temporal Impact Module, Side Information Model, and Periodic Impact Module, respectively. The ``Attention Weight'' refers to the corresponding attention score described in Equation \ref{equa:fuse}.
From the results in Table \ref{tab:module_compare}, we notice that the most effective combination incorporates all four modules, thereby demonstrating their collective contributions to the high-performance recommendations delivered by \shortname{}. According to the attention weight, the Popularity History Module holds the highest importance, followed by the Temporal Impact Module. 
% \hl{ This may be due to the less pronounced periodic fluctuations across different genres. (a bit of confused about the logic)}
Considering that our experiments on the \textit{Douban Movies} dataset only include attributes like categories, directors, and actors, we believe the influence of side information could potentially be enhanced if we integrate further details, such as user comments about the movies.
% Since our experiment only includes genres, directors, and actors on the \textit{Douban Movies} dataset, we think the impact of side information could potentially be enhanced if we incorporate additional information such as user comments on the movies.
% Additionally, our experiment only includes genres, directors, and actors on the \textit{Douban Movies} dataset. The impact of side information could potentially be enhanced if we incorporate additional information such as user comments on the movies.

\subsection{Performances of Item Popularity Prediction}

\begin{table}[]
\centering
\caption{Performance of two leading baseline models and their variants integrated with \shortname{} and \textit{Groundtruth} for Top-10 recommendation on \textit{Douban Movies}. \textit{Groundtruth} recommends the item with the highest actual popularity.}
\label{tab:groundtruth}
\resizebox{\columnwidth}{!}{ 
\begin{tabular}{ccccccc}
\\ \toprule
                       \multicolumn{2}{c}{Methods}            & Precision& Recall & HR & MRR & NDCG \\\midrule
\multicolumn{2}{c}{\shortname}              & 0.0208	& 0.1695	& 0.1994	& 0.0092	& 0.0955  \\
\multicolumn{2}{c}{\textit{Groundtruth}}       & 0.0360	& 0.3025	& 0.3371	&  0.0155	& 0.1666 \\\midrule
\multirow{3}{*}{STOSA} & Original    &0.0124	& 0.0870	& 0.1208	& 0.0062	& 0.0484 \\
                        & +\shortname        &0.0222	& 0.1730	& 0.2135	& 0.0080	& 0.0807 \\
                        & +\textit{Groundtruth} & 0.0365	& 0.3032	& 0.3399	& 0.0154	& 0.1644 \\\midrule
\multirow{3}{*}{ICLRec}  & Original    &0.0135	& 0.0961	& 0.1236	& 0.0054	& 0.0456\\
                        & +\shortname        & 0.0208	& 0.1695	& 0.1994	& 0.0094	& 0.0970 \\
                        & +\textit{Groundtruth} & 0.0376	& 0.3153	& 0.3567 & 0.0165	& 0.1752 \\ \bottomrule
\end{tabular}
}
\end{table}

\begin{figure*}
    \centering
    \subfloat[Distribution of items based on their predicted popularity scores and their actual (Ground Truth) popularity. The color of each data point indicates the number of items at that point.]{\includegraphics[width=.68\columnwidth]{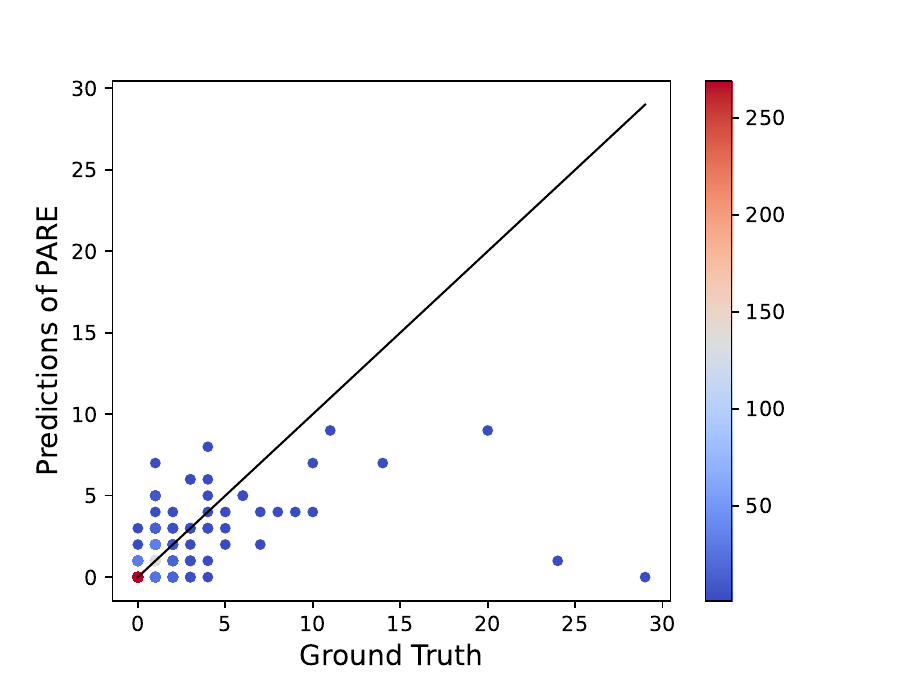}\label{fig:visul}}
    \hfill
    \subfloat[HR@10 comparison of 3 baseline models when integrated with \shortname{} through different $\beta$ on \textit{Douban Movies}.]{\includegraphics[width=.68\columnwidth]{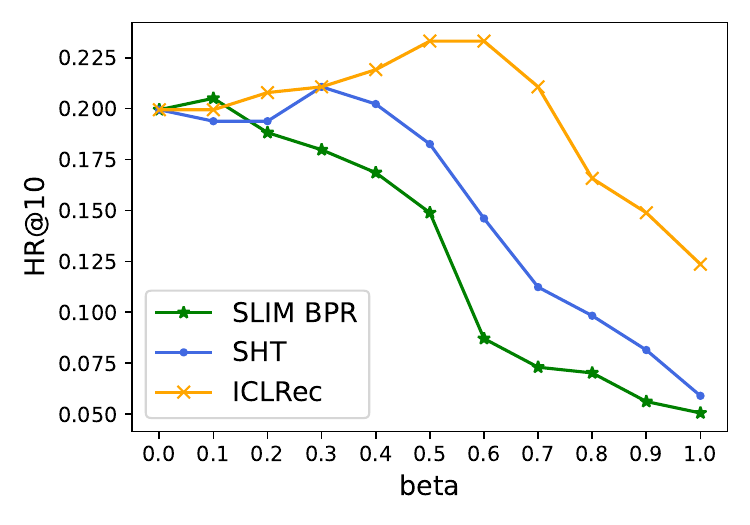}\label{fig:ablation_beta}}
    \hfill
    \subfloat[The number of overlapping items between the prediction list of \shortname{} and ICLRec on three datasets at varies Top-N recommendations.]{\includegraphics[width=.68\columnwidth]{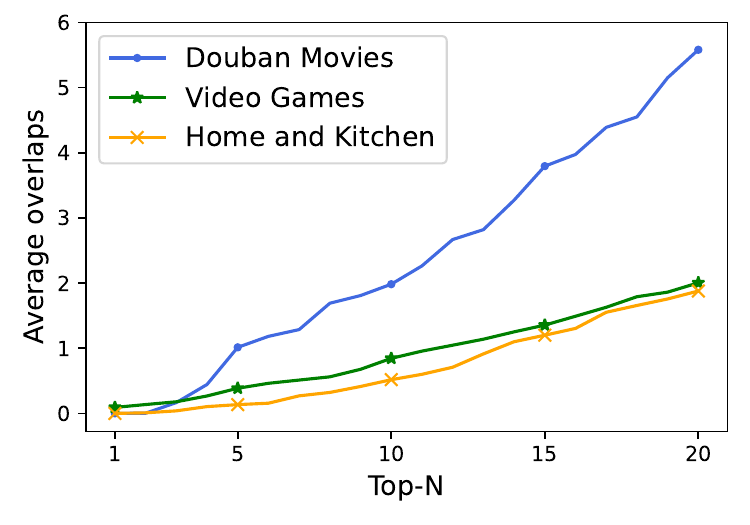}\label{fig:overlap_pop_model}}
    % \hfill
    % \subfloat[Distribution of items based on their predicted popularity scores and their actual (Ground Truth) popularity. The color of each data point indicates the number of items at that point.]{\includegraphics[width=.68\columnwidth]{fig/visualization.pdf}\label{fig:visul}}
    \caption{Ablation Study}
    % \caption{(a)Performances comparison of 3 baseline models when integrated with \shortname{} through different $\beta$ in \textit{Douban Movies}. (b)The number of overlapping items between the prediction list of \shortname{} and \hl{ICLRec} in three datasets at varies Top-N values. (c)Visualization for \shortname{} predictions and groundtruth item popularity. The predicted values are rounded to improve visibility. \hl{need check the caption format and content}}
    %\caption{Experimental figures on exploring the effectiveness of PARE as a complementary component (\ref{fig:ablation_beta}, \ref{fig:overlap_pop_model}), and analysing the accuracy of PARE in predicting item popularity (\ref{fig:visul}). }
    \label{fig:experiment_all}
\end{figure*}

In this experiment, we first analyze the accuracy of \shortname{} in predicting item popularity. 
Figure \ref{fig:visul} presents a visualization of the number of items as plotted against the predicted popularity and the actual popularity score from the test set of \textit{Douban Movies}. Alongside this, we also display a randomly selected equivalent number of items that are absent from the test set for comparison. We made the following observations. Firstly, for all items lacking interactions within the test set, the predicted popularity of these items does not exceed 3, highlighting the reasonable accuracy of \shortname{} when predicting for negative samples. 
Second, the predicted popularity of a significant portion of positive samples aligns closely with, or falls within 5 units of, the actual popularity, barring a few outliers.
According to these experimental results, \shortname{} can accurately predict item popularity to a certain extent.

Then, we aim to understand the upper-bound performance of non-personalized recommendation methods that solely rely on item popularity. To this end, we compare the performance of the top 10 recommendations generated by \shortname{} model with those recommending the item having the highest ground-truth popularity, represented as \textit{Groundtruth}, on \textit{Douban Movies}. 
Moreover, We evaluate the variants of two leading baseline models (i.e., STOSA and ICLRec), including the original approach, and versions integrated with \shortname{} and \textit{Groundtruth}. As illustrated in Table \ref{tab:groundtruth}, \textit{Groundtruth} surpasses \shortname{} by 78.43\% and 74.47\% in Recall@10 and NDCG@10, respectively, indicating potential room for improvement. Upon comparing the three variants of STOSA and ICLRec, it's evident that the original method lags behind in performance, while the version integrated with \textit{Groundtruth} significantly outperforms the other two variants. These insights further underscore the efficacy of taking item popularity into account when making recommendations.

\subsection{Effectiveness of \shortname{} as a Complementary Component}
To evaluate the effectiveness of \shortname{} when used as a complementary component to existing recommendation methods, we perform two ablation studies. First, as shown in Figure \ref{fig:ablation_beta}, we compare the HR@10 on \textit{Douban Movies} of varying $\beta$ values in Equation \ref{equa:integration} when \shortname{} is integrated with the best-performing recommendation baseline methods from traditional, non-sequential, and sequential methods, these being SLIM BPR, SHT, and ICLRec, respectively.
We observed that the integrated model with ICLRec exhibits superior performance when $\beta=0.6$. For SHT and SLIM BPR, the optimal performances were achieved at $\beta=0.3$ and $\beta=0.1$. As illustrated in Table \ref{tab:baseline_5metric}, ICLRec outperformed the other two baselines, followed by SHT and SLIM BPR.
These findings suggest that when the personalized recommendation model is not as strong, a lower $\beta$ value allows the integrated model to perform better, likely due to the increased emphasis on item popularity.

% We found that when $\beta=0.6$, the model integrated with ICLRec demonstrates superior performance. For SHT and ItemKNN, the optimal $\beta$ values is 0.3.
% As shown in Table \ref{tab:baseline_5metric}, ICLRec performs the best among the three baselines.
% % , followed by \hl{SHT and ItemKNN.} 
% Based on these findings, it can be concluded that in scenarios where the personalized recommendation model is less powerful, a higher $\beta$ value leads the integrated model to achieve superior performance. This is attributable to a greater emphasis on item popularity.

Then, we analyze the number of items that overlap between the recommendation list generated by \shortname{} and existing recommendation methods across three datasets. As depicted in Figure \ref{fig:overlap_pop_model}, we observe that the quantity of overlapping items increases approximately linearly as the number of items in the recommendation list grows. However, the absolute count of overlapping items remains relatively small, with approximately 2 items on \textit{Video Games} and \textit{Home and Kitchen} when recommending 20 items. This observation corroborates our assumption that due to the distinct approach of \shortname{} which doesn't rely on historical user-item interactions, it would yield a smaller overlap in the recommendation list compared to existing methods based on collaborative filtering.

In summary, these ablation studies show the benefits of integrating \shortname{} into existing recommendation models. Besides, these findings emphasize that when the original recommendation algorithm is deficient in effectiveness, the performance exhibits a more significant improvement upon integration with \shortname{}.

\section{Conclusion}
In conclusion, this paper has shed light on the critical influence of temporal fluctuations in item popularity for recommender systems. We identified that most existing recommendation methods focus on understanding users' personalized preferences through historical interactions, thereby often neglecting the dynamic shifts in item popularity. Addressing this gap, we propose \longname{} (\shortname{}), a non-personalized recommendation method by predicting the items likely to gain the highest popularity.

Our comprehensive experiments demonstrated \shortname{}'s capacity to compete with sophisticated state-of-the-art recommendation methods. Importantly, we found that \shortname{} can enhance the existing recommendation methods when incorporated as a complementary component. Given its simplicity, \shortname{} offers considerable practical utility for industrial applications and serves as a valuable baseline for future research in recommender systems.

\section{Acknowledgement}
This research is supported, in part, by Alibaba Group through Alibaba Innovative Research (AIR) Program and Alibaba-NTU Singapore Joint Research Institute (JRI), Nanyang Technological University, Singapore.

\newpage
\bibliographystyle{ACM-Reference-Format}
\bibliography{sample-base}

% \appendix

% \section{MovieLens Dataset Analysis}

% \begin{figure}
% 	\centering
% 	\includegraphics[width=0.9\columnwidth]{fig/movielens.pdf}
% 	\caption{Number of interactions in each month in MovieLens-1M dataset.}
% 	\label{fig:movielens}
% \end{figure}

\end{document}